\documentstyle[prl,aps,twocolumn,graphicx]{revtex}
\begin{document}
\draft

\draft \twocolumn[ \hsize\textwidth\columnwidth\hsize\csname
@twocolumnfalse\endcsname
 \title {Macroscopic Manifestation of Microscopic Entropy Production:
Space-Dependent Intermittence}
\author{Paolo Grigolini$^{1,2,3}$, Riccardo Mannella$^{3}$, Luigi 
Palatella$^{3}$}
\address{$^{1}$Center for Nonlinear Science, University of North Texas,\\
P.O. Box 305370, Denton, Texas 76203-5370}
\address{$^{2}$Istituto di Biofisica CNR, Area della Ricerca di Pisa,
Via Alfieri 1, San Cataldo 56010 Ghezzano-Pisa,  Italy}
\address{$^{3}$Dipartimento di Fisica dell'Universit\`{a} di Pisa and
INFM,
Piazza Torricelli 2, 56127 Pisa, Italy}
\date{\today}

\maketitle \widetext
\begin{abstract}
We study a spatial diffusion process generated 
by velocity fluctuations of intermittent nature.  
We note that intermittence reduces the entropy production 
rate while enhancing the diffusion strength. We study a case 
of space-dependent intermittence and prove it to result 
in a deviation from  uniform distribution. 
This macroscopic 
effect can be used to measure the relative value of the trajectory entropy.
\end{abstract}

 ] \narrowtext

The problem of detecting the macroscopic effects of microscopic 
randomness has been an active field of research for several 
years. The literature is so extended that we limit ourselves
to only a few sample 
papers\cite{sample1,sample2,sample3,sample4,sample5}. 
The problem is hard and controversial. It is possible to derive
the ordinary diffusion equation, often used in these studies,
as the asymptotic limit of a dynamic prescription with no irreversibility
ingredients\cite{howard}, thereby raising the problem itself 
of where to locate the source of entropy production. This source is often
identified with the random nature of the microscopic trajectories, 
which, at least in principle, is rigorously defined by the 
Kolmogorov-Sinai (KS) entropy $h_{KS}$\cite{dorfman}. 
Gaspard and Nicolis\cite{sample1} pointed out that the transport 
coefficients of open systems, and notably the diffusion coefficient, 
can be expressed as the difference between the positive Lyapunov 
exponent and the KS entropy, thereby implying that the macroscopic 
manifestation of the trajectory randomness has to be looked for in 
non-equilibrium processes, in line with the tenets of irreversible 
thermodynamics\cite{hystory}.

Here we discuss a different way to the macroscopic manifestation 
of the trajectory entropy, based on the 
diffusion of a space variable $x$ collecting the intermittent 
fluctuations of a random velocity $\xi$. We show
that in the case where the velocity intermittence is space dependent,
the equilibrium reached by the system is the same as that 
produced by a temperature gradient, even if the 
kinetic energy of a single trajectory remains rigorously constant 
throughout the whole observation process. Thus, a ``paradoxical'' 
equilibrium distribution shows up, reminiscent of the dynamical 
Maxwell's Demon effect, recently discussed by Zaslavsky\cite{zaslavsky}.
This effect is anomalous but is not unphysical, and we show that it can 
be used to establish the relative value of trajectory entropy.

The velocity $\xi$ is a dichotomous variable, namely, with
only two 
values, $+W$ or $- W$, and we set these values in sequence as
follows. We randomly  
draw integer numbers $l$, interpreted as time lengths in units of $\tau$. 
We use a random number generator, assumed to draw  
with uniform probability the numbers 
$s$ of the interval $[0,1]$. After any drawing we associate 
the selected number $s$ to  $y = 
- (1/\lambda) \ln s$. 
Then we consider only the integer part of it, $l = [y]+1$.
The probability of large $l$'s is so 
high that we can adopt a continuous rather than a discrete picture, 
for both the interval lengths, $l$, and the time associated to them, 
$t = l\tau$. 
The probability of drawing the time $t$ is proportional to:
\begin{equation}
\psi(t) = \frac{\lambda}{\tau} \exp \left(- \frac{t}{\tau} \lambda
\right).
\label{integers}
\end{equation}
Adopting the jargon of the authors of Ref.\cite{geisel}, who used 
intermittently chaotic maps to derive faster than normal 
diffusion, 
we refer to $\psi(t)$ as the \emph{waiting time 
distribution of the laminar phase}. This means that the $i$-th drawing
selects the integer number $l(i)$, which defines a time interval $l(i) 
\tau$ corresponding to the velocity $\xi$ maintaining the value $+W 
(-W)$, without changing sign. At the $(i+1)$-th step, when the length 
$l(i+1)$ is selected, the velocity $\xi$ changes sign and keeps the 
new value $-W(+W)$ for the 
whole interval of time $l(i+1) \tau$. There is no 
uncertainty associated with the choice of the velocity sign. 
Randomness is involved only at the moment of drawing the number $s$. 
We call $H(E)$ the entropy increase associated with this drawing. This
quantity is assumed to be the same for any drawing, and, consequently, 
as we shall see, the Maxwell's Demon effect is independent of it.
It is evident that the sequence of symbols $\xi(t)$ 
generated by this procedure is characterized by the steady rate of 
entropy increase
\begin{equation}
    h_{E} = H(E)\frac{\lambda}{\tau},
    \label{externalentropy}
    \end{equation}
since $\tau/\lambda$ is the mean time spent by the velocity in 
the laminar phase. We refer to this entropy as External (E) 
entropy to point out that it does not coincide with the KS 
entropy.
   
What about the connection between the KS and the E-entropy? This 
question can be answered considering the asymmetric Bernoulli map 
studied by Dorfmann \emph{et al.} \cite{etal}. We describe this map, 
changing the  notations of 
Ref.\cite{etal}, as follows. We consider the variable $z$, moving in 
the interval $[0,1]$, with the condition of folding back into this 
interval the portion of $z$ exceeding the value $z= 1$, and with 
the following dynamic recipe:
\begin{equation}
z_{n+1}= z_{n}/q,
\label{laminarregion}
\end{equation}
for $ 0 \leq z_{n}< q$, and
\begin{equation}
z_{n+1} = (z_{n} - q)/p,
\label{chaoticregion}
\end{equation}
for $q \leq z_{n}< 1$. We assume $q = 1 -p$ and we consider this map 
in the case where $p <<1$. This means that the KS entropy of this 
map\cite{etal},
\begin{equation}
h_{KS} = (1-p) \ln \left(\frac{1}{1-p}\right)
+ p \ln \left(\frac{1}{p}\right),
\label{etal}
\end{equation}
becomes (using a time step $\tau$, not necessarily equal to $1$)
\begin{equation}
h_{KS} = \frac{p}{\tau} \left[1 + \ln \left(\frac{1}{p}\right)\right].
\label{pzero}
\end{equation}
Let us assign the symbol $+W$ to the ``laminar'' region of 
Eq. (\ref{laminarregion}) and the symbol $-W$ to the ``chaotic'' 
region of Eq. (\ref{chaoticregion}). The resulting sequence is 
apparently different from that produced by the prescription 
earlier adopted to generate the fluctuations of the variable 
$\xi$. However, the numerical calculation proves that in the 
limiting case $p << 1$ the two 
kinds of symbolic sequences yield the same KS entropy\cite{marco}. 
The reader can easily convince him/herself that the two KS 
entropies are identical by noticing that the only element of 
randomness concerns the prediction about the time duration of 
the laminar phase. The distribution of the laminar phase waiting times
is given by Eq.(\ref{integers}) with $\lambda =  
p/(1-p)$.
The KS entropy can be written in a form reminiscent of that of 
the E-entropy of Eq. (\ref{externalentropy}),
\begin{equation}
h_{KS} = H(KS)\frac{\lambda}{\tau},
\label{KSentropyasEentropy}
\end{equation}
provided that the uncertainty $H$ is now defined as
\begin{equation}
H(KS) \equiv \left(\frac{\lambda}{1 + \lambda}\right)\left[1 + \ln 
\left(\frac{1 + \lambda}{ \lambda}\right)\right].
\label{KSuncertainty}
\end{equation}
The source of uncertainty is the same as that mirrored by the 
E-entropy, namely, the random drawing of numbers of the interval 
$[0,q]$, which, in fact, in the limiting case of $p \rightarrow 0$
coincides with the interval $[0,1]$. However, the KS entropy reflects 
also the fact that the random selection of these numbers is operated 
by the chaotic region of Eq.(\ref{chaoticregion}). The E-entropy, on 
the contrary, interprets the sporadic randomness as produced by an external 
source (hence the term \emph{external} used 
to denote it) with a strength independent of $\lambda$.

The diffusing variable is the space variable $x(t)$ that in the 
case when $t/\tau >>1$ can be related to the velocity $\xi(t)$ by 
\begin{equation}
x(t) = \int_{0}^{t} \xi(t') dt' + x(0).
\label{trajectory}
\end{equation}
In the special case we are considering, the variable $\xi(t)$ is 
dichotomous, thereby implying the property 
\begin{eqnarray}
\langle \xi(t_{1})\xi(t_{2})
\ldots\xi(t_{2n-1})\xi(t_{2n}) \rangle = ~~~~~~~~~~
\nonumber \\ 
\langle\xi(t_{1})\xi(t_{2})\rangle 
\ldots\langle\xi(t_{2n-1})\xi(t_{2n})\rangle,
\label{dichotomousrandomness}
\end{eqnarray}
the correlation functions with odd numbers of times vanishing as a 
consequence of the assumption made that no bias exists,
namely, $\langle\xi(t)\rangle = 0$.
In this condition, it is straightforward to prove that for $1 \leq n$
\begin{eqnarray}
\langle x^{2n}(t)\rangle = (2n)! 
W^{2n}
\int_{0}^{t}dt_{1}\int_{0}^{t_{1}}dt_{2}\ldots  \times ~~~~~~ \nonumber \\
\int_{0}^{t_{2n-1}}dt_{2n}
\Phi_{\xi}(t_{1}-t_{2})\ldots\Phi_{\xi}(t_{2n-1}-t_{2n}),
\label{secondmoment}
\end{eqnarray}
where $\Phi_{\xi}(t)$ denotes the normalized correlation 
function of $\xi$ ($\Phi_{\xi}(0) = 1$).
It is also straightforward to show that the Liouville-like 
equation, generating all the moments of 
Eq. (\ref{secondmoment}), and thus properly 
describing this diffusion process, is:
\begin{equation}
\frac{\partial}{\partial t} \sigma(x,t) 
= W^{2}\int_{0}^{t} \Phi_{\xi}(t-t') 
\frac{\partial^{2}}{\partial x^{2}}\sigma(x, t')dt'.
\label{exactliouville}
\end{equation}
On the other hand, although this equation is expected to be an 
exact picture of the dichotomous process under study, it can also 
be derived, via a projection procedure, from a standard Liouville 
picture applied to an additional set of variables,
${\bf \Psi}$, responsible for the random-like behavior of the variable 
$\xi$\cite{allegro}, as well as to  $x$, $\xi$. We note that in the
asymptotic time limit 
this equation, under the assumption alone that the correlation 
function $\Phi_{\xi}(t)$ is integrable, yields the ordinary diffusion 
equation
\begin{equation}
\frac{\partial}{\partial t} \sigma(x,t)= D\frac{\partial^{2}}{\partial 
x^{2}}\sigma(x,t).
\label{ordinary}
\end{equation}
Using the proper connection between $\psi(t)$ and 
$\Phi_{\xi}(t)$\cite{allegro,zumofen,note}, we
express the diffusion coefficient as
\begin{equation}
D = W^{2}\frac{\tau}{2 \lambda} .
\label{diffusioncoefficient}
\end{equation}

It is embarrassing, in our opinion, that the same result as that
produced by the random microscopic trajectory here under study, 
the ordinary diffusion equation of Eq. (\ref{ordinary}), can be 
derived from a perspective with no explicit use of microscopic randomness 
\cite{howard}, hence keeping alive the long standing debate on the 
origin of thermodynamics and statistical mechanics that, 
according to Boltzmann\cite{lebowitz} rests on the key action 
of infinitely many degrees of freedom. This is probably the 
strength of statistical mechanics, whose successes rest on fundamental 
equations, such as the Fick's law\cite{howard}, independent  of
the philosophical 
perspective adopted for their derivation. 
Here, we prove that microscopic randomness can produce
ostensible macroscopic effects. This raises the interesting issue
of how to derive these effects from within the deterministic 
perspective\cite{howard,lebowitz}.
   
\begin{figure}[htb]
\centering
\includegraphics*[angle=-90, width=3.4in]{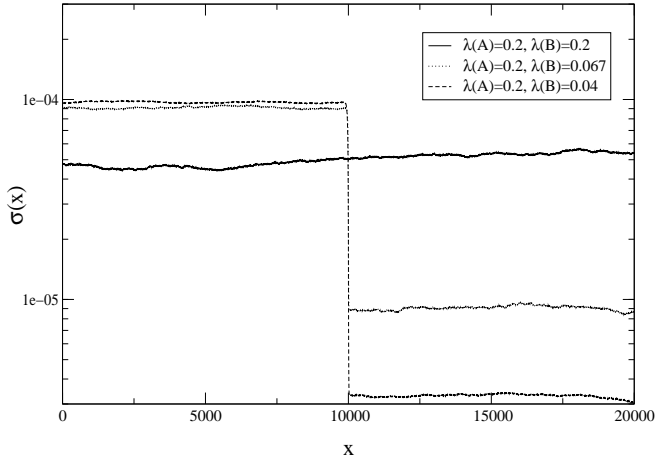}
\caption[]{Typical equilibrium distributions.}
\end{figure}

We have seen that  $1/\lambda$
is the time dilatation intensity. The parameter $\lambda$, consequently, measures the 
intermittence-induced reduction of entropy production.  
To prove how a form of anomalous statistical mechanics can 
emerge out of this property, let us consider the case 
where the particle moves in the interval $[0,L]$. In the case of 
ordinary statistical mechanics, the system of diffusing particles,
after a fast transient process
reaches a condition of uniform equilibrium distribution, thereby
opening a temporary window for the observation of entropy production, 
which, in fact, from the pioneer times of Ref.\cite{hystory} to these 
days\cite{sample3}, is done in out of equilibrium conditions.
Here we draw benefits from the intermittent nature of the microscopic 
process, controlled by  $\lambda$. 
We call $A$ and $B$ the left and right portion of the container, and 
we assign different values of $\lambda$ to them, while leaving the 
kinetic energy constant. The time dilatation in $A$ is smaller than in 
 $B$, namely, $\lambda(A) > \lambda(B)$. At the moment of the
$j$-th drawing, the number $s(j)$ generates either 
$y(j) = - (1/\lambda(A)) \ln s(j)$  
or  $y(j) = - (1/\lambda(B)) \ln s(j)$, according to whether the 
particle is in $A$ or $B$ at the end of the earlier laminar phase.
At $x = 0$ and $x = L$, the velocity $\xi$ changes signs, so as to 
mimic the elastic collision with the walls, and keeps it till to 
exhaust the sojourn time selected by the last drawing.  
According to 
Eq. (\ref{diffusioncoefficient}), the diffusion process in $A$  is 
slower than in $B$, thereby making, as shown 
by Fig. 1, $\sigma(A) >\sigma(B)$, 
where $\sigma(A)$ and $\sigma(B)$ are the particle 
densities in $A$ and $B$, respectively. On the basis of the 
earlier theoretical arguments, it is evident that this Maxwell's 
Demon effect\cite{zaslavsky} is determined by the fact that the 
trajectory entropy in $A$ is larger than in $B$.

\begin{figure}[htb]
\centering 
\includegraphics*[angle=-90, width=3.4in]{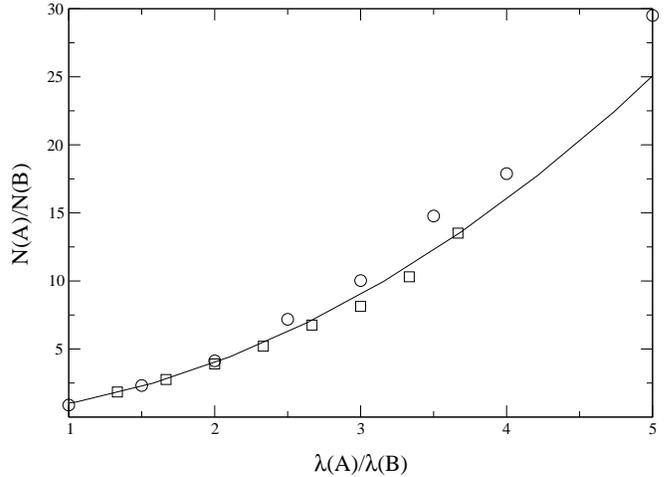}
\caption[]{Ratio between equilibrium populations in the two half
spaces, as function of the lifetimes ratio. Solid line is theory,
circles are simulation results based on the integration of a 
single random trajectory, squares are simulation results based on the
long time distribution obtained evolving an initially flat distribution.}
\end{figure}

In fact, the numerical experiment proves that (Fig. 2) 
\begin{equation}
N(A)/ N(B) =  [h_{E}(A)/h_{E}(B)]^{2} ,                
\label{paradox}
\end{equation}
where   $N(A)$ and $N(B)$ denote the number of particles in $A$ and 
$B$, respectively, and 
$h_{E}(A)$ and $h_{E}(B)$ are the corresponding external entropies. 
This numerical result can be given a plausible theoretical 
explanation. We note that the traditional diffusion equation of 
Eq. (\ref{ordinary}) is a proper description of the diffusion 
process in a time scale larger than $ \tau/(2\lambda)$. 
This observation yields 
the following balance equation, between
$\sigma(A)$ and $\sigma(B)$, 
\begin{equation}
D(A) \sigma(A) \Delta(A\rightarrow B) = D(B) \sigma(B) 
\Delta(B\rightarrow A)
\label{crucialbalance},
\end{equation}
where $D(A)$ and $D(B)$ denote the diffusion coefficients of $A$ 
and $B$, respectively, evaluated according to the prescription 
of Eq. (\ref{diffusioncoefficient}). The rationale for this key balance 
is that, according 
to the theory of the first passage time\cite{gardiner}, the 
diffusion coefficients $D(A)$ and $D(B)$ can be interpreted as 
the rate of the processes of diffusion from the walls at $x =0$ and 
$x=L$, respectively, to the border between the two portions. 
Thus the number of particles entering the portion $B$, per unit 
of time, $N(A\rightarrow B)$, is proportional to $D(A)$. However, the larger 
is the memory
of the initial condition, the larger is the number of particles
moving from $A$ to $B$. Thus, $N(A\rightarrow B)$ must be proportional 
also to a spatial window of size $\Delta(A\rightarrow B)$. Similar 
arguments are used for $N(B\rightarrow A)$, and
Eq. (\ref{crucialbalance}) follows. 
It is natural to assume $\Delta(A\rightarrow B)$ and $\Delta(B\rightarrow A)$
to be proportional to $W\tau/\lambda(A)$
and $W\tau/\lambda(B)$, respectively. This assumption,
Eq. (\ref{diffusioncoefficient}) and Eq. (\ref{crucialbalance}) yield 
a theoretical prediction coinciding with Eq. (\ref{paradox}).

In conclusion, $h_{E}$, as well as $h_{KS}$,  is a form of trajectory
entropy, difficult to  
detect from within a macroscopic equilibrium condition. We have proved
that the space  
dependent intermittence yields unusual, but not unphysical, 
macroscopic equilibrium properties that can be used to measure the
relative value  
of the E-entropy, and, hence, through the relations between the 
E-entropy and the KS entropy, established in this letter, to derive 
also information on the KS entropy. The Maxwell's Demon effect of 
this letter is similar to that of Zaslavsky, but not identical to it. 
As shown by Zaslavsly\cite{book}, in his model the left container, 
corresponding to the portion $A$, is a Cassini billiard and the 
right container, corresponding to the portion $B$, is a Sinai
billiard. The larger density in $A$  
is not caused by a slower diffusion, but by a more extended sojourn 
time in the state of regular collisions with the walls (the laminar 
state of the Zaslavsky model). However, we are convinced that by properly 
adapting to that case the perspective established in this letter, it should be 
possible to express also the Zaslavsky version of the Maxwell's Demon effect 
as a measure 
of the relative value of the trajectory entropy. In that case the 
randomness associated to the collision between particle and scatterer 
is made sporadic by long sojourns in states corresponding to regular 
collisions with the walls of the container. 

Finally, we like to address the connection between the dynamic systems
of this letter and the \emph{bona fide} turbulent process\cite{uriel}, 
and also the related problem of the thermodynamic nature of a 
L\'{e}vy gas. The inverse power law character of the waiting time 
distribution is essential to produce the macroscopic manifestation of 
microscopic intermittence, under the form of anomalous diffusion. The 
Maxwell's Demon effect, on the contrary, does not require the waiting 
time distribution to have an inverse power law nature. As shown in this 
letter, this form of macroscopic manifestation of intermittence only 
requires the random bursts to occur with a space-dependent frequency. 
In fact, the time dilatation strength used in this letter  is not constant, 
and $1/\lambda (A) < 1/\lambda(B)$. On the other hand, the exponential 
condition used in this letter is not necessary: 
The important property 
behind the Maxwell's Demon effect is that the mean sojourn time in the 
laminar region changes with moving from one to the other portion of the 
container, including the possibility of being infinite in one of the 
two portions. Thus, our approach can be easily extended to the case of 
inverse power law: It is enough to take\cite{luigi} 
$y=T[1/(1-s)^{1/(\mu-1)}-1]$. The mean waiting time of the resulting 
distribution is not affected by the transition from the Gauss to the 
L\'{e}vy statistics, occurring at $\mu = 3$, and it keeps a finite 
value in the whole interval $[2,3]$, but at $\mu =2$, where it 
diverges. This divergence signals a transition to a non-stationary 
condition, incompatible with the dynamic derivation of L\'{e}vy  
statistics\cite{bettin}. This phase transition makes the E-entropy, and 
with it the KS entropy\cite{gaspard}, vanish  for $\mu \leq 2$. Thus, 
this letter suggests that the temperature of a L\'{e}vy process becomes 
infinite at the onset of this transition.


\begin{references}
\bibitem{sample1} P. Gaspard and G. Nicolis, Phys. Rev. Lett. 
{\bf 65}, 1693 (1990).

\bibitem{sample2} V. Latora and M. Baranger, Phys. Rev. Lett. {\bf 82},
520 (1999).

\bibitem{sample3} T. Gilbert, J. R. Dorfman and P. Gaspard, 
Phys. Rev. Lett. {\bf 85},
1606 (2000).

\bibitem{sample4} W. H. Zurek and J. P. Paz, Phys. Rev. Lett. {\bf 72},
2508 (1994).

\bibitem{sample5} A. K. Pattanayak, Phys. Rev. Lett. {\bf 83}, 
4526 (1999). 

\bibitem{howard} M. H. Lee, Phys. Rev. Lett. 
{\bf 85}, 2422 (2000).

\bibitem{dorfman}J. R. Dorfman, \emph{An Introduction to Chaos in 
Nonequilibrium Statistical Mechanics}, Cambridge Lecture Notes in 
Physics (Cambridge University Press, Cambridge, 1999).

\bibitem{hystory} S. R. de Groot and P. Mazur, \emph{Nonequilibrium 
Thermodynamics} (Dover, New York, 1984). 

\bibitem{zaslavsky} G. M.  Zaslavsky, Physics Today {\bf 52}(8), 39 (1999). 

\bibitem{geisel} T. Geisel, J. Nierwetberg and A. Zacherl, Phys. Rev. 
Lett. {\bf 54}, 616 (1985).

\bibitem{etal} J. R. Dorfman, M. H. Ernst and D. Jacobs, 
J. Stat. Phys. {\bf 81}, 497 (1995). 

\bibitem{marco} P. Gri\-go\-li\-ni, M. G. Pa\-la and L. Pa\-la\-tel\-la,
in cond-mat/0007323.

\bibitem{allegro} P. Allegrini, P. Grigolini and B. J. West, Phys. Rev. 
E {\bf 54}, 4760 (1996).

\bibitem{zumofen} G. Zumofen and J. Klafter, Phys. Rev. E {\bf 47}, 851 
(1993). 

\bibitem{note} Using the prescription of Ref.\cite{zumofen}, it is 
straigthforward to replace the process with alternated velocity signs, 
studied in this letter, with an equivalent process, with randomly 
selected velocity signs, of which Ref.\cite{allegro} affords an exact 
treatment.


\bibitem{lebowitz} J. L. Lebowitz, Physica A {\bf 263}, 516 (1999).

\bibitem{gardiner} C. W. Gardiner, \emph{Handbook of Stochastic Methods, 
Second Edition} (Springer, Stuttgart, 1985).

\bibitem{book} G. M. Zaslavsky, \emph{Physics of Chaos in Hamiltonian 
Systems} (Imperial College Press, London, 1998).

\bibitem{uriel} U. Frisch, 
\emph{Turbulence, the legacy of A. N. Kolmogorov} (Cambridge University 
Press, Cambridge, 1995).

\bibitem{luigi} M. Buiatti, P. Grigolini and  L. Palatella, Physica A {\bf 
268}, 214 (1999).

\bibitem{bettin} R. Bettin, R. Mannella, B. J. West and 
P. Grigolini,  Phys. Rev. E {\bf 51}, 212 (1995).

\bibitem{gaspard} P. Gaspard and X.-J. Wang, Proc. 
Natl. Acad. Sci. USA {\bf 85}, 4591 (1988).

\end{references}
\end{document}